# Communication scientifique : Pour le meilleur et pour le PEER

Laurent Romary, INRIA-Gemo & HUB-IDSL

## Introduction

Les fondements de la communication scientifique ont profondément évolué au cours des 15 dernières années. En réponse à la crise des revues (*serial crisis*) et conscient des nouvelles potentialités technologiques, nombre de chercheurs, bibliothécaires ou décideurs ont cherché à mettre en œuvre des moyens de diffuser largement et si possible librement l'ensemble du patrimoine des publications scientifiques. Ceci s'est traduit notamment par la prolifération au niveau international d'archives dites institutionnelles, permettant à tout chercheur de déposer ses productions pour une diffusion immédiate en ligne.

Cependant, la situation globale est loin d'être claire, et au delà du caractère très fragmenté que présente le paysage des archives de publications[1], il est parfois difficile d'appréhender l'usage effectif qu'en font les chercheurs, ainsi que l'impact que cet usage, potentiellement croissant, peut avoir sur le paysage de l'édition scientifique traditionnelle. Une meilleure compréhension de ces phénomènes, à la fois sur le plan technique et sociologique, doit permettre entre autre d'améliorer les modes de collaboration avec les éditeurs, mais aussi de définir des schémas directeurs pour les politiques d'information scientifique des organismes de recherche.

Fruit d'un débat initialement très animé au niveau européen, le projet PEER, vise à apporter des éléments de réponse dans le cadre d'un espace de collaboration technique et conceptuel entre éditeurs et chercheurs.

## Un contexte politique agité

Les réflexions ayant mené à la définition du projet PEER remontent au début de l'année 2007, quand, dans le contexte de l'annonce par la Commission européenne de « recommandations concernant l'évolution technique et économique des marchés de l'édition scientifique en Europe »[2], les communautés des chercheurs et des éditeurs se sont lancés dans une bataille de tranchées à coup, respectivement, de pétitions[3] et de communiqués de presse[4], plaçant *de facto* la Commission européenne en position d'arbitre. De part et d'autre, la conscience est forte que des avancées politiques majeures sont en train de se préparer. Les chercheurs, et les organisations dont ils font partie, anticipent la possibilité d'une véritable obligation de dépôt pour les publications en Europe, ce que veulent freiner à tout prix les éditeurs.

---

[1] Cf. Romary et Armbruster, Beyond institutional repositories, à paraître dans Int. J. of Digital Library Systems, 2010 (http://hal.archives-ouvertes.fr/hal-00399881/fr/)

[2] Cf. l'étude commanditée par la commission : http://ec.europa.eu/research/science-society/pdf/scientific-publication-study_en.pdf

L'idée est alors pour la Commission, dans le contexte notamment des réflexions menées par le groupe d'experts sur les bibliothèques numériques sous les auspices de la DG INFSO[5], de réunir autour d'une même table des représentant d'institutions scientifiques et d'éditeurs pour définir des actions communes visant notamment à offrir une transition souple vers différents modèles de diffusion libre de l'information scientifique. Très tôt émerge l'idée d'un *observatoire de l'édition scientifique* dont la mission serait de suivre les évolutions du domaine et d'identifier, à l'aide de différents indicateurs (production scientifique, usages, modèles économiques sous-jacents) les éventuels changements dans ses grands équilibres.

Le projet PEER[6], financé dans le cadre du programme eContent+ est une première étape pour développer conjointement un tel observatoire, en se focalisant sur ce qui touche aux archives de publications.

## PEER : une plate-forme de communication les éditeurs et la communauté scientifique

Le projet PEER repose sur un consortium de cinq partenaires représentant à la fois les éditeurs scientifiques[7] (STM, association internationale des éditeurs des domaines scientifique, technique et médical), les organismes de recherche (Max Planck Society, INRIA[8]), les Universités (Bibliothèque de l'Université de Göttingen) et les agences de recherches (ESF – European Science Foundation). Le projet collabore par ailleurs étroitement avec la fondation SURF[9], l'université de Bielefeld et l'initiative européenne Driver[10].

---

[3] http://www.ec-petition.eu/

[4] Plusieurs documents sont disponibles dans les archives de l'association STM (http://www.stm-assoc.org)

[5] http://ec.europa.eu/information_society/activities/digital_libraries/index_en.htm

[6] http://www.peerproject.eu/

[7] Les éditeurs suivants sont directement impliqués dans PEER : BMJ Publishing Group; Cambridge University Press; EDP Sciences; Elsevier; IOP Publishing; Nature Publishing Group; Oxford University Press; Portland Press; Sage Publications; Springer; Taylor & Francis Group; Wiley-Blackwell.

[8] www.inria.fr, voir aussi HAL-INRIA : http://hal.inria.fr/

L'objectif principal du projet PEER est d'observer les effets d'un dépôt systématique et à grande échelle des manuscrits-auteurs après évaluation par les pairs, à savoir la phase 2 du cycle de vie d'un article scientifique (phase 1 : manuscrit soumis ; phase 3 : version finale de l'éditeur). Les questionnements envisagés tournent autour des aspects suivants :

- Impact du dépôt systématique sur la viabilité des journaux (impact sur les abonnements et/ou les téléchargements sur le site de l'éditeur) ;
- Impact sur la diffusion de l'article scientifique (accroissement cumulés des accès et/ou des citations) ;
- Impact sur l'écologie globale des productions de recherche (e.g. transferts de soumission d'un journal vers un autre) ;
- Détermination des facteurs favorisant le dépôt par les auteurs et évaluation des coûts correspondants ;
- Conceptions de modèles permettant de faire coexister l'édition scientifique « traditionnelle » avec l'auto-archivage.

De fait, PEER est conçu comme un véritable espace de communication entre scientifiques et éditeurs. A un premier niveau, il s'agit de mieux comprendre les perspectives propres à chaque groupe vis-à-vis de la question de l'« *open acces* » en général et de la place des archives de publication en particulier. A un deuxième niveau, des réponses communes sont envisagées de façon à réduire les craintes ou les défiances de part et d'autre. Typiquement, la contribution des éditeurs au dépôt systématique, d'une part, et l'identification de critères de confiance (gestion des périodes mobiles d'embargo, fournitures de statistiques d'accès) pour les archives de publication, d'autre part, visent à mieux articuler les archives de publications aux processus éditoriaux traditionnels.

Enfin, et ce n'est pas nécessairement l'aspect le plus négligeable, le projet PEER est l'occasion de définir des directives précises sur l'échange tant des métadonnées que des textes intégraux. Le contexte technique actuel est en effet très fragmenté avec notamment l'absence de stratégie claire de normalisation de part et d'autre.

---

[9] http://www.surffoundation.nl/en/Pages/default.aspx

[10] http://www.driver-repository.eu/

## Le corpus – la méthode

L'ensemble des travaux du projet PEER repose sur l'identification d'un corpus de plus de 300 journaux scientifiques fournis par les éditeurs participants, et répondant aux critères suivants :

- 20% du contenu des journaux doit être européen (déterminé par l'auteur-correspondant) ;

- La qualité, reflétée notamment par le facteur d'impact, doit être particulièrement bonne ;

- Autant que possible, les domaines concernés doivent être relativement larges ;

- Les périodes d'embargos, telles que définies par les éditeurs, doivent être suffisamment courtes (12 mois ou moins) pour permettre des observations fiables au cours du projet.

Le corpus de journaux est divisé en deux groupes : l'un pour lequel les éditeurs déposent directement les méta-données et le manuscrit-auteur, et l'autre où un courrier aux auteurs les incite à déposer par eux-mêmes.

Le dépôt proprement dit, qu'il s'agisse d'un dépôt-éditeur ou d'un dépôt auteur, s'effectue par le biais d'un portail (le « PEER Depot ») qui gère la normalisation des données et redirige les informations sur les archives de publication qui participent au projet.

## Les domaines de recherche

Les différentes activités de recherche envisagées au sein du projet ne sont pas conduites directement pas les membres du consortium, mais sous-traitées, après appel d'offre, à des équipes indépendantes[11]. Les axes de recherche, initiés depuis avril 2009, portent sur les aspects suivants :

- Recherche sur les comportements : l'équipe du département des sciences de l'information et du LISU à l'université de Loughborough (UK) est chargée d'identifier les comportements des auteurs et des usagers d'archives de publication ;

- Recherche sur les usages : le groupe CIBER de l'UCL (University College London) est chargée de définir les origines et les conditions d'usage des articles déposés dans les archives de publication et de définir des indicateurs facilitant le suivi de tels usages ;

- Recherche sur les modèles économiques : l'appel d'offre lancé en septembre 2009, porte sur la détermination de l'ensemble des coûts (éditeurs et archives) associés aux différents modes de dépôt et d'accès envisagés, et la mise en perspective de ces coûts avec les coûts actuels de publication d'articles scientifiques.

---

[11] Une partie non négligeable du management lié au projet consiste, comme on peut s'en douter, à gérer le plus finement possible les éventuels conflits d'intérêt.

## La plate-forme technologique

Comme nous l'avons vu, le projet PEER repose sur un dépôt centralisé couplé à un réseau parfaitement déterminé d'archives de publication (« trusted repositories »). Ces archives ont accepté de suivre un certain nombre de règles techniques précises (modèles de données, interface SWORD de dépôt) et éditoriales (fourniture de statistiques communes normalisées, gestion fine des affiliations) garantissant un traitement uniforme du dépôt des manuscrits-auteurs.

De même, les éditeurs acceptent de se conformer à un certain nombre de contraintes liés notamment à la fourniture d'un jeu minimal de métadonnées, et permettant le suivi précis des informations liées à chaque article. Par contre il a été décidé d'accepter tous les formats (XML) des différents éditeurs, afin d'avoir une idée de la complexité des opérations de normalisation à réaliser.

Le « PEER Depot » joue un rôle particulier en ce qu'il reçoit toutes les informations primaires, en vérifie les contenus puis transfère une version normalisé vers les différentes archives. Il permet en fait aux éditeurs et aux archives d'avoir un interlocuteur technique unique qui assure la cohérence et la complétude des opérations de dépôt. L'une des activités importantes du « PEER Depot » a ainsi été de définir un format conforme aux directives de la TEI (Text Encoding Initiative[12]) permettant de reprendre le contenu intégral des articles ainsi que les métadonnées correspondantes.

La spécification du processus de dépôt est entièrement documentée dans un rapport de référence disponible en ligne :

*Guidelines for publishers and repository managers on deposit, assisted deposit and self-archiving*[13]

## Bilan

Un premier bilan à mi-projet permet d'ores et déjà de constater que le projet PEER a véritablement conduit à une amélioration importante du dialogue entre éditeurs et chercheurs sur les attendus réciproques vis-à-vis des archives de publications. Non seulement de nouveaux éditeurs ont souhaité rejoindre le projet en cours de route, mais le climat de confiance autour de la mise en place, tant de la plate forme de dépôt, que des actions de recherche, a permis de valider les processus de transfert

---

[12] http://www.tei-c.org

[13] http://www.peerproject.eu/fileadmin/media/reports/D3_1_Guidelines_v8.3_20090528.Final.pdf

d'information, ce qui n'était pas nécessairement acquis au moment du lancement du projet.

Sur le plan technologique, la normalisation des formats[14] nécessitera probablement des développements qui vont au delà du projet PEER lui-même. Il apparait en effet nécessaire de définir un vrai cadre de standardisation pour les articles scientifiques qui favorise à la fois un accès facilité aux contenus et une perspective d'archivage pérenne de ceux-ci. Des formats *ad hoc*, du type de la DTD NLM (National Library of Medecine), spécifiés sans véritable stratégie de qualité technique, de maintenance ou de documentation (ce qui caractérise un standard), et utilisés de façon très variable dans les milieux de l'édition nuisent en fait à la large diffusion des publications scientifiques.

Enfin, il est important de signaler que les partenaires académiques du projet ne sont pas naïfs. Il est clair que le milieu de l'édition cherche à maintenir un marché dont le rendement, au regard d'autres domaines d'activités du secteur de l'information, est particulièrement élevé. Il reste que dans un contexte de forte pression politique au niveau européen, avec la perspective d'une obligation intégrale de dépôt pour les publications associées aux projets du huitième programme cadre, il est important de mieux connaître les modèles de collaboration possibles entre éditeurs et académiques, ainsi que les schémas viables de déploiement d'archives de publication qui peuvent y être associés.

---

14 cf. Holmes, M., & Romary, L. (2009). Encoding models for scholarly literature. In S. Kapidakis (Ed.), *Publishing and digital libraries: Legal and organizational issues.* http://hal.archives-ouvertes.fr/hal-00390966/fr/